\documentclass[useAMS, usenatbib]{mnras}
\usepackage{aas_macros}
\usepackage{url}
\usepackage{bm}
\usepackage{graphicx}
\usepackage{amsmath}

\title[A HAWKI survey of late T dwarfs]{Robust detection of quasi-periodic variability: A HAWKI mini survey of late T dwarfs}

\author[Stuart Littlefair]{S.\,P.\ Littlefair$^{1}$, B.\ Burningham$^{2}$, Ch.\ Helling$^{3}$\\
$^{1}$Dept of Physics and Astronomy, University of Sheffield, Sheffield, S3 7RH, UK \\
$^{2}$Centre for Astrophysics Research, Science and Technology Research Institute, University of Hertfordshire, Hatfield AL10 9AB, UK \\
$^{3}$SUPA, School of Physics and Astronomy, University of St Andrews, St Andrews KY16 9AA, UK\\
}

\begin{document}



\maketitle

\label{firstpage}

\begin{abstract}
We present HAWK-I $J$-band light curves of five late-type T dwarfs (T6.5--T7.5) with a typical duration of four hours, and investigate the evidence for quasi-periodic photometric variability on intra-night timescales. Our photometry reaches precisions in the range 7--20 mmag, after removing instrumental systematics that correlate with sky background, seeing and airmass. Based upon a Lomb-Scargle periodogram analysis, the latest object in the sample - ULAS J2321 (T7.5) - appears to show quasi-periodic variability with a period of 1.64 hours and an amplitude of 3 mmag.

Given the low amplitude of variability and presence of systematics in our lightcurves, we discuss a Bayesian approach to robustly determine if quasi-periodic variability is present in a lightcurve affected by red noise. Using this approach, we conclude that the evidence for quasi-periodic variability in ULAS J2321 is not significant. As a result, we suggest that studies which identify quasi-periodic variables using the false alarm probability from a Lomb-Scargle periodogram are likely to over-estimate the number of variable objects, even if field stars are used to set a higher false alarm probability threshold. Instead we argue that a hybrid approach combining a false alarm probability cut, followed by Bayesian model selection, is necessary for robust identification of quasi-periodic variability in lightcurves with red noise.
\end{abstract}


\bibliographystyle{mnras}

\section{Introduction} 
Clouds form in the photospheres of brown dwarfs of most spectral types. L-type dwarfs have thick clouds of iron and silicates \cite[e.g][]{tsuji96, allard01, marley02, burrows06}. Around the L/T transition these clouds either drop below the photosphere or break up into patches; as a result, the early T dwarfs are thought to be relatively cloud free \cite[e.g.][]{ackerman01, burgasser02}. In the late-T dwarfs condensates of alkali salts and sulfides are believed to form; models which include this opacity provide improved fits to the near- and mid-infrared colors of late-T dwarfs \citep{morley12}. However, it is worth noting that \cite{line15} find via hierarchical Bayesian model selection that (grey) clouds are not justified by the spectra of late-T dwarfs.

Observations of photometric variability in brown dwarfs provide a method for probing the presence and structure of condensate clouds. If the cloud deck is longitudinally heterogeneous, photometric variability will occur as parts of the cloud deck rotate in and out of view.  Over the past 15 years, numerous surveys have attempted to detect photometric variability in brown dwarfs \cite[see][and references within for a review]{metchev15}. Major breakthroughs occurred when  \cite{artigau09} detected strong ($\Delta J = 50$ mmag) quasi-periodic variability in the T2.5 dwarf SIMP J013656.5+093347 and \cite{radigan12} found even stronger ($\Delta J=260$ mmag) quasi-periodic variability in the T1.5 dwarf 2MASS J21392676+0220226. These results triggered large surveys for near-infrared variability which showed that photometric variability is common, but typically of low amplitude. \cite{radigan14a} surveyed 57 L4--T9 dwarfs for variability; they found that 16\% of their targets showed photometric variability in the $J$-band. Both the amplitude and frequency of variability is enhanced near the L/T transition, with 39\% of L/T transition objects showing evidence for variability. Outside the transition typical amplitudes are 0.6--1.6\%. The small amplitude of variability, and challenges of near-infrared photometry, mean that care must be taken to distinguish astrophysical variability from instrumental systematics. For example, the BAM survey \citep{wilson14} found 14 variables amongst 69 surveyed objects, many with large amplitudes (\textgreater 2\%) and at all spectral types. However a re-analysis of their data by \cite{radigan14b} found that much of this variability was attributable to instrumental systematics. Space-based surveys can attain higher precision. Whilst they also suffer from instrumental systematics, in many cases these are well understood and can be removed from the data, to some extent. \cite{metchev15} surveyed 44 L3--T8 dwarfs with the Spitzer space telescope. They find that low-amplitude variability is exceedingly common, with 80\% of L dwarfs varying with amplitudes $\ge0.2$\% and 40\% of T dwarfs varying with amplitudes $\ge0.4$\%.

These surveys have largely focused on spectral types of mid-T and earlier, typically including one or two later T dwarfs. Surveys for variability in late T dwarfs have the potential to test predictions of increased cloud opacity due to suflide and alkali salt condensates; the formation of new cloud layers may be accompanied by an increase in the frequency and amplitude of variability at later spectral types \citep{morley14}. The existing data show that variability exists amongst late T dwarfs; \cite{metchev15} finds clear variability in the T6 dwarf 2MASS J22282889-4310262 with an amplitude of 5\% and a periodicity of 1.4 hours in the  [3.6] band. \cite{rajan15} carried out small scale survey of three late-T dwarfs and one Y-dwarf. They report a detection of 13\% variability with a periodicity of 3 hours in the T8.5 dwarf WISEP J045853.89+643452.9AB in one epoch, however this was not seen in a repeat visit. Recently, \cite{cushing16} find periodic variability with a period of 8.5 hours and semi-amplitude of 3.5\% in the Y dwarf WISE J140518.39+553421.3. Whilst it is clear that variability exists amongst the late T and early Y dwarfs, more data is needed for good statistical estimates of its amplitude and frequency of occurrence.

As a caveat to the above discussion, we note that heterogeneous cloud coverage is not the only potential cause of variability in brown dwarfs. Magnetic spots are probably ruled out due to the neutrality of the atmosphere \citep{mohanty02}, although lightning discharge may increase the electron density in the atmosphere \citep{bailey14}. Another source of photometric variability may be  temperature variations due to dynamical perturbations \citep{robinson14, zhang14}. Recently, photometric variability in a late M-dwarf has also been attributed to auroras \citep{hallinan15}. The photometric variability in this case is thought to be caused by non-thermal ionisation of the atmosphere produced by the impact of the auroral electron current. Similar photometric variability may be present in brown dwarfs of all spectral types; the coherent, pulsed radio emission associated with auroras is seen down to spectral types as late as T6.5 \citep{route12, route16}. Indeed, it is worth pointing out that near-infrared variability and radio emission seem to be closely related, and that the prototype variable T-dwarf SIMP J013656.5+093347 has been detected in the radio \citep{kao16}. Hybrid variability mechanisms are also plausible, with either temperature fluctuations or non-thermal ionisation influencing the heterogeneity of the condensate cloud deck. Whilst it is likely that the majority of brown dwarf variability is due to condensate clouds, time resolved spectrophotometric observations are required to disentangle the contribution of other mechanisms.

Motivated by the relative sparsity of data on the variability of late T dwarfs, we undertook a mini-survey using HAWK-I on the VLT. We selected targets for observation from the ${\sim}$150 brown dwarfs with spectral types between T4 and T9 in the UKIDSS Large Area Survey \cite[][and references therein]{burningham13,burningham10}. Many of these objects have multi-epoch observations as a result of their initial survey photometry and subsequent follow-up. We found that of the 95 objects with 2 epochs of $J$-band observations, 15\% have discrepancies greater than 3$\sigma$ in their photometry. To select the most likely candidates for photometric follow-up, we selected the 5 objects with spectral types close to T7 that have discrepant $Y$- and $J$-band photometry regardless of aperture size. These objects are listed in table~\ref{tab:obs}.

The observations are described in section~\ref{sec:observations}. The results are presented in section~\ref{sec:results}. In section~\ref{sec:modeling} we describe our use of Bayesian model comparison to test for the presence of periodic variability. The results are discussed in section~\ref{sec:discussion}.

\section{Observations and Data Reduction}
\label{sec:observations}

\begin{table*}
\caption{Journal of observations.}
\label{tab:obs}
\centering
\begin{tabular}{llcccccc}
Name & Alias & SpT & UTC Start & UTC End & Seeing (\arcsec) &  Airmass & Photometric? \\
\hline
WISE J234026.61-074508.1 & WISE 2340-0745 & T7 & 2012-10-17 23:40 & 2012-10-18 04:31 & 0.5--1.1 & 1.3 --1.7 & N \\
ULAS J015024.37+135924.0 & ULAS J0150+1359 & T7.5 & 2012-10-18 04:39 & 2012-10-18 08:06 & 0.5--0.7 & 1.3--2.2 & Y \\
WISE J234841.10-102844.1 & WISE 2348-1028 & T7 & 2012-10-18 23:35 & 2012-10-19 04:21 & 0.3--0.5 & 1.0--1.4 & Y \\
ULAS J230601.02+130225.0 & ULAS J2306+1302 & T6.5 & 2012-10-19 23:34 & 2012-10-20 03:41 & 0.4--0.6 & 1.3--1.6 & Y \\
ULAS J232123.79+135454.9 & ULAS J2321+1354 & T7.5 & 2012-10-20 23:31 & 2012-10-21 04:18 & 0.3--0.5 & 1.3--1.7 & Y \\
\end{tabular}
\end{table*}

On the nights between 17th Oct 2012 and 20th Oct 2012 we observed our target stars using HAWKI \citep{kissler-patig08} on UT4. Observing conditions are shown in table~\ref{tab:obs}. All objects were placed near the centre of chip \#3. Three objects (ULAS J2306, ULAS J2321 and WISE J2345) were observed in a 5-position offset pattern. 60 second exposures were taken in each offset position. We refer to this as pattern A. Two objects (ULAS J0150 and WISE J2340) were taken in a different offset pattern, with the aim of increasing time resolution. In this pattern 6 seperate 10 second exposures were taken at each offset position. We refer to this as pattern B.

The individual images were flat-fielded, dark- and background-subtracted using v.1.8.12 of the HAWKI pipeline recipes. The sky-background subraction is done in two steps. An initial pass is run on the first complete offset pattern (5 frames for pattern A, 30 frames for pattern B). In this pass a background image is produced from a median of the data and the data are background subtracted, aligned and co-added to produce a master frame in which objects are detected. The object mask is then used in a second pass to prevent bright objects from contaminating the background.  For pattern A the background for an observation is the median of 7 frames either side of the observation. When computing the background value for each pixel, the two lowest and highest frames are rejected. For pattern B, 42 frames either side of the observation were used, and the five lowest and highest frames are rejected.

After background subtraction, the frames are aligned using the locations of bright stars and aperture photometry is carried out using the {\sc ultracam} data reduction software\footnote{\url{https://github.com/trmrsh/cpp-ultracam}}. To achieve the best photometry we experimented with a range of aperture sizes and extraction methods. The best results were obtained with standard aperture photometry using apertures which were scaled with the seeing; aperture sizes ranged from 1 to 1.3 times the full-width-half-maximum (FWHM) measured in each frame. 

\begin{figure*}
\begin{center}
\includegraphics[scale=0.45,clip=true,trim=0 0 0 0]{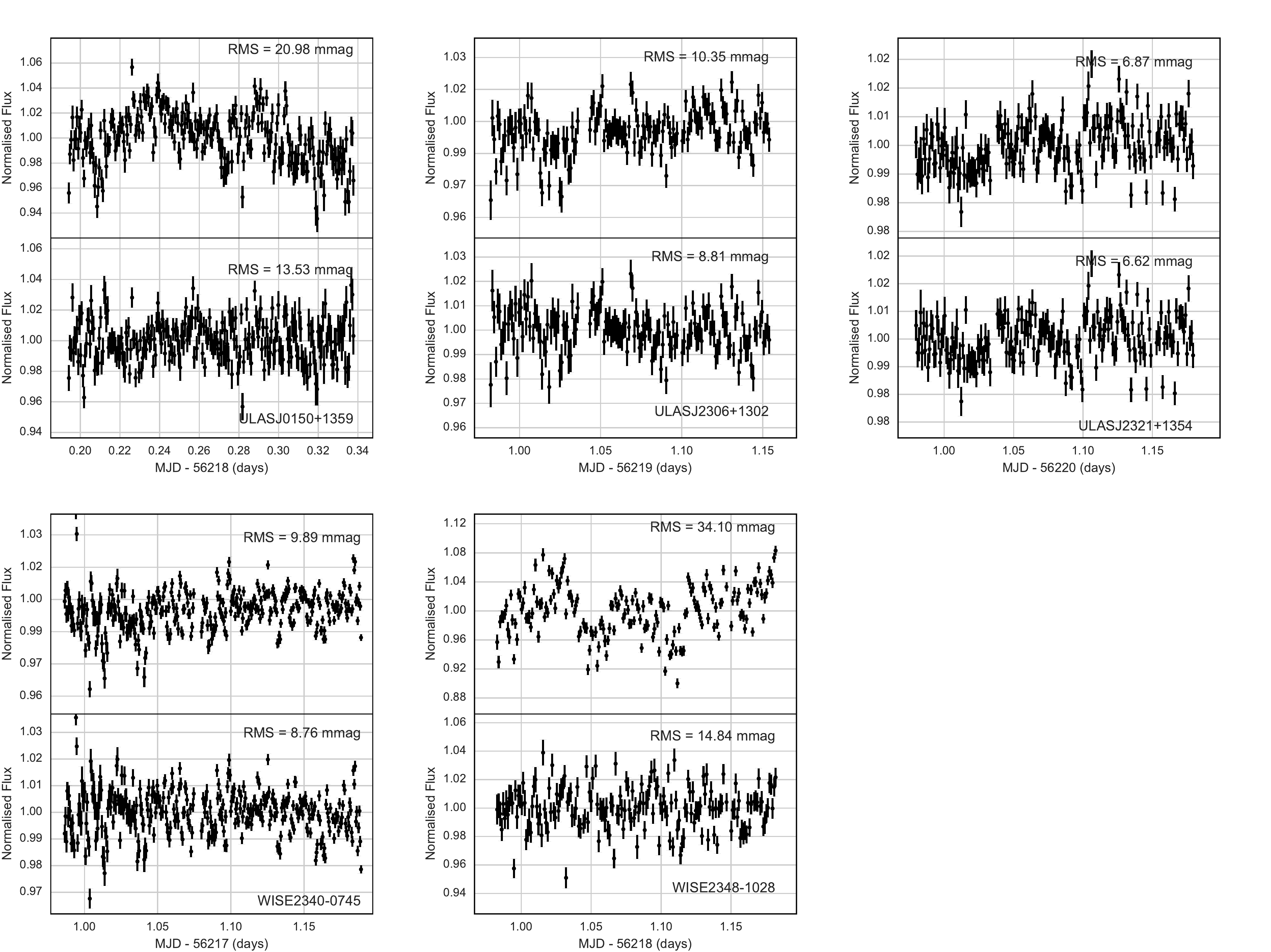}
\caption{HAWKI $J$-band lightcurves for all stars observed. Each sub-panel shows the data before (top) and after (bottom) removing systematics (see text for details). The lightcurves have been normalised by dividing a constant fit to the data.}
\label{fig:lcs}
\end{center}
\end{figure*}

We extracted photometry for our targets and all bright stars located on the same chip as our target. The lightcurves of the bright stars were combined using a weighted mean to produce a reference lightcurve which was used to correct the lightcurve of our target star. We experimented with different choices of comparison stars to give the lightcurve with the smallest root-mean-square deviations from a constant flux. The results are shown in figure~\ref{fig:lcs}. 

\section{Results}
\label{sec:results}
All of the brown dwarf lightcurves in figure~\ref{fig:lcs} show smooth trends and higher frequency variability. The same patterns, with similar amplitudes, are seen in the lightcurves of our comparison stars. This strongly suggests that the variability is instrumental, rather than intrinsic in origin. 

We can attempt to remove some of this instrumental variability by finding and removing correlations with airmass, sky background and seeing.  Our data consists of a vector of times \mathbfit{t}, fluxes \mathbfit{y} and uncertainties $\bm{\sigma}$. We also have measurements of the sky background \mathbfit{s}, the average FWHM \mathbfit{f} and the airmass \mathbfit{X}. For each object, the lower panel in figure~\ref{fig:lcs} shows the data after dividing by $a\mathbfit{s} + b\mathbfit{f} + c\mathbfit{X}$. The coefficients $(a, b, c)$ are chosen to minimise the variance of the decorrelated data  \mathbfit{y'}. This process greatly reduces the variability apparent in the data, supporting our belief that much or all of this variability is due to instrumental systematics. 

One possible source of instrumental variability is second-order extinction; in the J-band late T-dwarfs emit predominantly at wavelengths that are relatively unaffected by telluric absorption. Differences in telluric absorption between our targets and the references stars can introduce spurious variability. To some extent, this is accounted for by decorrelating agains airmass above. We also repeated the analysis above, adding a linear dependence on the relative humidity, which we use as a proxy for the line-of-sight water column. Adding this term does not significantly improve the removal of systematics.

\begin{figure*}
\begin{center}
\includegraphics[scale=0.45,clip=true,trim=120 0 0 0]{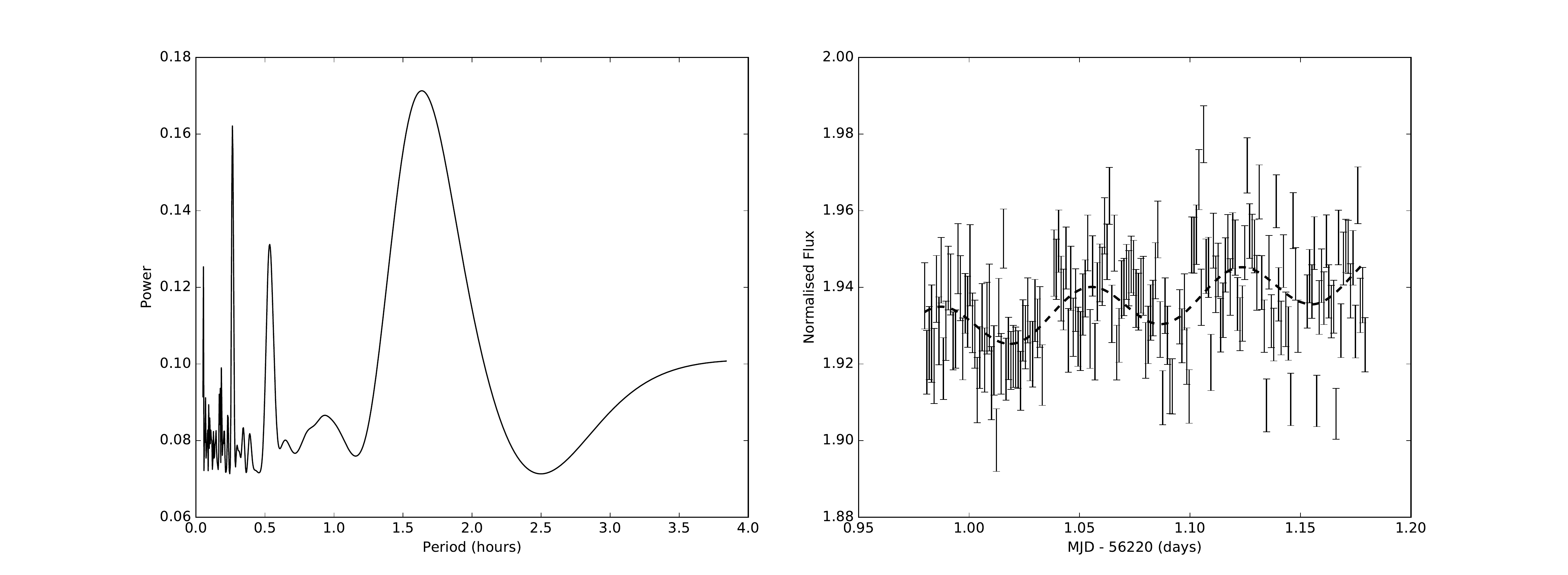}
\caption{{\bf Left: } The generalised Lomb-Scargle periodogram of the $J$-band lightcurve of ULAS~J2321+1354. {\bf Right: }  The $J$-band lightcurve of ULAS~J2321+1354 before decorrelation. The best fitting sinusoid and linear trend, with a period fixed at a value of 1.64 hours as suggested by the periodogram, is shown as a dashed line. The amplitude of the best fitting sinusoid is 3 mmag.}
\label{fig:lomb}
\end{center}
\end{figure*}

For four of the five brown dwarfs, the residual variability is stochastic and has a typical amplitude of 10 mmag, or 1\%. If intrinsic to the brown dwarfs, this may be caused by a turbulent atmosphere driving stochastic variability. However, we do not believe this variability is real, instead attributing it to instrumental systematics that have not been removed by the decorrelation process described above. The exception is the lightcurve of ULAS~J2321+1354. The raw lightcurve appears to show a gradual linear trend and hints of periodic variability. Both the linear trend and the periodic variability survive the decorrelation process. We calculated a generalised Lomb-Scargle periodogram \citep{zechmeister09}, which accounts for a floating mean and observational errors. The periodogram was calculated using {\sc gatspy} \citep{vdp15}, and shows a strong peak at a period of 1.64 hours. The periodogram, and best-fitting sinusoid are shown in figure~\ref{fig:lomb}. We calculated the false-alarm probability (FAP) for the strongest peak following \cite{zechmeister09}. The FAP is given by
\begin{equation}
{\rm FAP} = 1 - [1-{\rm Prob}(z > z_{0})]^{M},
\end{equation}
where ${\rm Prob}(z > z_{0})$ is the probability that a periodogram power $z$ can exceed the highest peak found $z_{0}$, under the assumption that the data are Gaussian noise. $M$ is the number of independent frequencies. Using this formula we found FAP = 0.5\%. Using the same method, the FAP for our comparison star lightcurves indicates that none of our comparison stars show significant evidence for periodicity; the lowest FAP amongst our comparison lightcurves was 10\%. 

Whilst the Lomb-Scargle periodogram of ULAS~J2321+1354 shows evidence of periodic variability with a period of 1.64 hours, one cannot take the FAP at face value. Estimating the number of independent frequencies is difficult for unevenly spaced data. Monte-Carlo methods can overcome this \cite[e.g][]{cumming99}, but are computationally expensive. In 100,000 trials where we calculated the periodogram of Gaussian noise with the same time sampling as the ULAS~J2321+1354 lightcurve, 7 had a power greater than that seen in the actual periodogram, leading to an estimate of the false-alarm probability of ${\rm FAP}_{\rm MC} = 0.007$\%. 

Even Monte-Carlo methods cannot overcome a more serious issue in using the FAP to estimate the reality of an apparent periodic signal. Critically, the FAP gives the probability that a peak as large as the one observed will occur in the periodogram by chance, {\em under the assumption that the data are pure Gaussian noise}. Many astrophysical lightcurves do not satisfy this assumption. Instrumental systematics or astrophysical red noise mean that even in the absence of periodic signals, the lightcurve has clear temporal structure which is not consistent with pure Gaussian noise. This is clearly the case for our data, where instrumental systematics are present. As a result, we do not believe the FAP is an appropriate measure to judge the reality of the periodic signal in ULAS~J2321+1354. Instead, we describe below an alternative method to determine the reality of a periodic signal in data which also shows systematics or astrophysical red noise, based on computing the Bayesian evidence for competing models.

\section{Bayesian Model Fitting}
\label{sec:modeling}

Consider two models $\mathcal{M}_{A}$ and $\mathcal{M}_{B}$. Given our dataset \mathbfit{d} we can write the posterior odds ratio:
\begin{equation}
\frac{P(\mathcal{M}_{A} \mid  \mathbfit{d} )}{P(\mathcal{M}_{B} \mid  \mathbfit{d} )} = \frac{Pr(\mathcal{M}_{A})}{Pr(\mathcal{M}_{B})} 
\cdot \frac{ m(\mathbfit{d} \mid \mathcal{M}_{A} )}{ m(\mathbfit{d} \mid \mathcal{M}_{A} )}.
\end{equation}
$\frac{Pr(\mathcal{M}_{A})}{Pr(\mathcal{M}_{B})}$ is the prior odds ratio: we set this ratio to unity as we have no prior reason to prefer either model. The second term on the right-hand side of the equation is the ratio of marginal likelihoods, also known as the evidence ratio or Bayes' factor. The marginal likelihood $m$ of a dataset \mathbfit{d} given a model $\mathcal{M_{A}}$ with parameter set $\theta_{A}$ is given by:
\begin{equation}
m(\mathbfit{d} \mid \mathcal{M_{A}}) = \int{ \mathcal{L} (\mathbfit{d} \mid \mathcal{M_A}, \theta_{A}) \, Pr(\theta_{A} \mid  \mathcal{M_A}) \; d\theta_{A} },
\label{eq:like}
\end{equation}
where $\mathcal{L} (\mathbfit{d} \mid \mathcal{M_A}, \theta_{A})$ is the likelihood function for the data, given the model and a particular set of parameters.

Under this framework, we could determine if a periodic signal is real by computing the odds ratio between two models. Our first model, $\mathcal{M_A}$, would contain parameters to fit both the red noise in the dataset and a periodic component. The second model, $\mathcal{M_B}$, would only contain the red noise parameters. A Bayesian approach to finding periodic signals thus has two challenges; finding appropriate models for the dataset, and adopting a suitable algorithm to calculate the marginal likelihood of these models, given these data.

\subsection{Gaussian process models for red noise and periodicity}

Gaussian processes (GPs) are seeing increasing use to model systematics and red noise in astrophysical datasets. For a textbook introduction see \cite{rasmussen06}. \cite{roberts13} offers a clear explanation of how GPs can be used to represent time-series data, and examples of applications to a wide range of datasets. Our dataset \mathbfit{d} consists of $n$ points with times \mathbfit{t}, fluxes \mathbfit{y} and uncertainties $\bm{\sigma}$. A GP represents the dataset as a multivariate Gaussian distribution. The covariance matrix $\bm{K}$ describes the extent to which each pair of data correlate with each other. Pure white noise would be described by covariance matrix where only the diagonal elements were non-zero. Allowing every element of the covariance matrix to be fitted would require $n\times n$ parameters, so the problem is made tractable by adopting a kernel function $k(t_{i}, t_{j})$. The elements of the covariance matrix are then given by:
\begin{equation}
K_{ij} = \sigma_{i}^{2} \delta_{ij} + k(t_{i}, t_{j}).
\end{equation}

Various choices of kernel function exist; common choices for modelling red noise are the Mat{\'e}rn-3/2 (M32) function:
\begin{equation}
k(t_{i}, t_{j}) = a^{2} \left( 1 + \sqrt{ \frac{3r^{2}}{\tau^{2}} } \right) \exp{\left( - \sqrt{\frac{3r^2}{\tau^2}} \right)},
\end{equation}
and the squared-exponential (SE) function:
\begin{equation}
k(t_{i}, t_{j}) = a^{2} \exp {\left( -\frac{r^2}{2\tau^2} \right)},
\end{equation}
where $r^{2} = (t_{i} - t_{j})^{2}$. $a$ and $\tau$ are the parameters of the GP model; $a$ represents the typical amplitude of the variability due to red noise, and $\tau$ represents the timescale of variations. SE kernel functions give rather smooth variations, whereas M32 kernel functions are better suited to model rougher variations \citep{roberts13}.

The lightcurves shown in figure~\ref{fig:lcs} appear to show systematics on two timescales; there are short timescale systematics which have quite rough structure and smoother, long timescale systematics. Motivated by this observation, we adopt the following kernel function to model the systematics in our data:
\begin{multline}
k(t_{i}, t_{j}) = a_s^{2} \left( 1 + \sqrt{ \frac{3r^{2}}{\tau_s^{2}} } \right) \exp{\left( - \sqrt{\frac{3r^2}{\tau_s^2}} \right)} + \\
	a_l^{2} \exp {\left( -\frac{r^2}{2\tau_l^2} \right)},
\label{kfunc}
\end{multline}
where $a_{s}$ and $\tau_{s}$ are the amplitudes and timescales respectively of the short timescale variability and $a_{l}$ and $\tau_{l}$ are the amplitudes and timescales of the longer timescale trends. 

To calculate the marginal likelihood for a model we first need to be able to evaluate the likelihood for a specific set of parameters. Given a function $f_{\theta'}(\mathbfit{t})$ that has parameters $\theta'$ and represents the mean of the data, we can calculate the residuals, $ \mathbfit{r}  = \mathbfit{y} - f_{\theta'}(\mathbfit{t})$. The likelihood is given by \citep{rasmussen06}:
\begin{equation}
\ln \mathcal{L}(\mathbfit{d} \mid \theta) = -\frac{1}{2} \mathbfit{r}^{T} \mathbfit{K}^{-1} \mathbfit{r} - \frac{1}{2} \ln \det \mathbfit{K} - \frac{n}{2} \ln 2 \pi,
\end{equation}
where $\theta$ is the full set of parameters, including $\theta'$ and the parameters for the Gaussian process. We used the {\sc george}\footnote{\url{http://dan.iel.fm/george}} package \citep{ambikasaran14} to implement our GP kernels.

Finally then, we are ready to define two models we can compare to test for the presence of periodic variability. Our first model,  $\mathcal{M_A}$, consists of the kernel function given by equation~\ref{kfunc} and a sinusoidal mean function:
\begin{equation}
f_{\theta'}(\mathbfit{t}) = \mu + A \sin{ \left[ 2\pi \left( \frac{\mathbfit{t} - t_0}{P} \right) \right] },
\end{equation}
where $\mu$ is the mean level of the lightcurve, $A$ the amplitude of the sinusoidal signal, $t_0$ represents the phase of the sinusoid and $P$ is the rotational period of the brown dwarf. Our second model, $\mathcal{M_B}$, combines the same kernel function with a simple mean function with one parameter:
\begin{equation}
f_{\theta'}(\mathbfit{t}) = \mu.
\label{simple_mean}
\end{equation}

These two models would be sufficient to test for the presence of a non-evolving, purely sinusoidal signal. However, the rotational variability seen in many brown dwarfs to date is more complex than this \cite[e.g][]{metchev15, radigan12, artigau09} . Multiple surface features, and evolution of the surface features on timescales comparable to the rotational period, can cause the lightcurve to differ drastically from a simple sinusoid. We describe a third model, $\mathcal{M_C}$, which combines the simple mean function of equation~\ref{simple_mean} with a Gaussian process to represent quasi-periodic variability and short timescale systematics. The kernel function for this Gaussian process is given by:

\begin{multline}
k_{\rm rot}(t_{i}, t_{j}) = A^{2} \exp \left[ -\Gamma \sin^{2} \left( \frac{\pi |r|}{P} \right) - \frac{r^2}{\tau_{l}^{2}} \right] + \\
 a_s^{2} \left( 1 + \sqrt{ \frac{3r^{2}}{\tau_s^{2}} } \right) \exp{\left( - \sqrt{\frac{3r^2}{\tau_s^2}} \right)}.
\label{kfunc_rot}
\end{multline}
The first term in this kernel function is the quasi-periodic function used to model rotational variability by \cite{aigrain16},  \cite{vanderburg15} and \cite{haywood14}. This consists of an exponential-sine-squared kernel (ESn2) multiplied by an squared-exponential (SE) kernel. The addition of a M32 kernel models the short timescale systematics. $P$ is the rotational period and $A$ is the amplitude of quasi-periodic variability. The parameter $\Gamma$ is a ``roughness'' parameter; it controls the regularity of the quasi-periodic lightcurve. Small values of $\Gamma$ will give sinusoidal lightcurves, whereas large values will produce more irregular variations with larger contributions from harmonics. $\tau_{l}$ is the evolutionary timescale of the periodic variations. 

\begin{figure*}
\begin{center}
\includegraphics[scale=0.5,clip=true,trim=80 20 0 0]{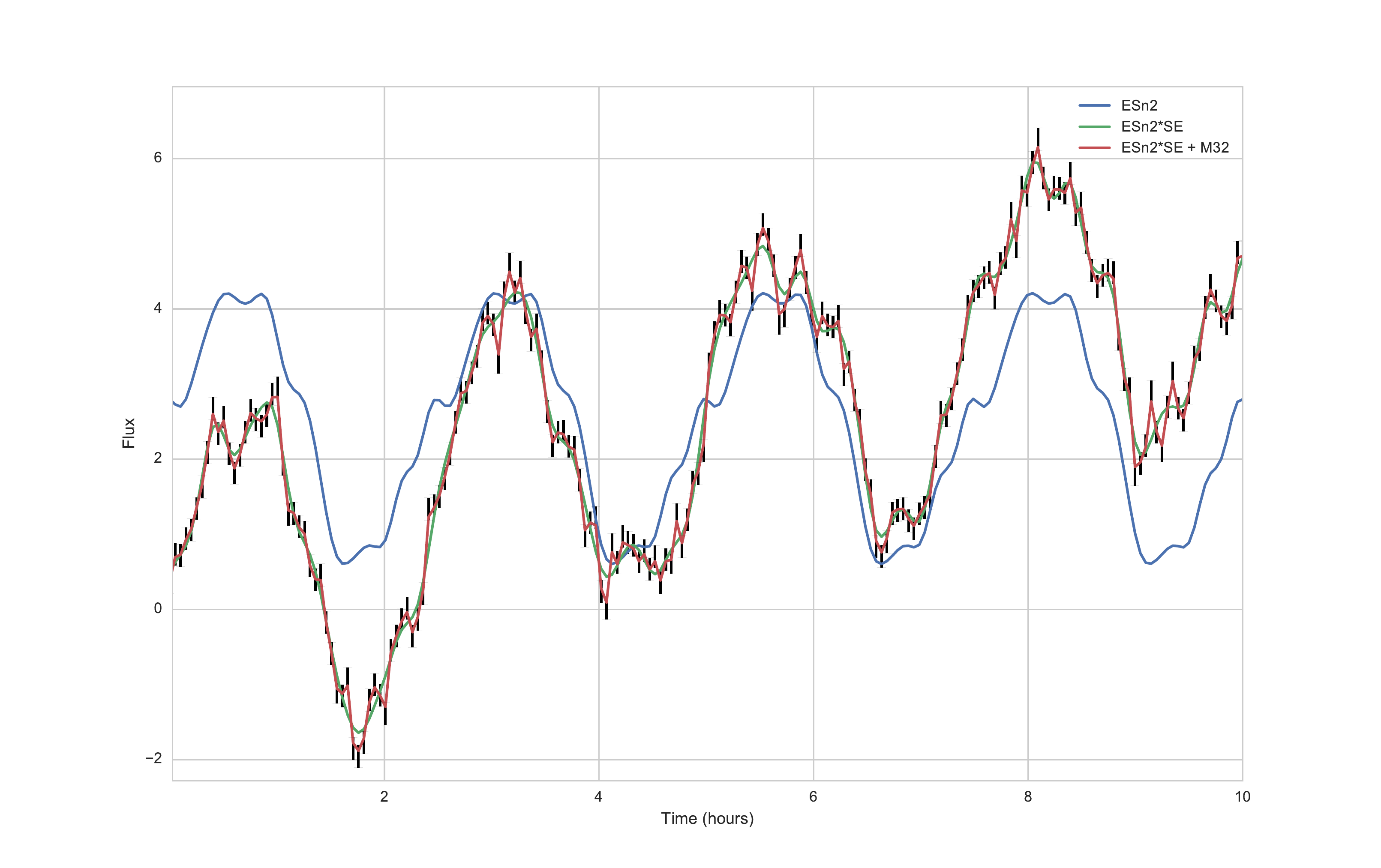}
\caption{Representation of quasi-periodic variability with additional systematics by a Gaussian process. Fake data (black error bars) is created by adding a sinusoidal term, a long term linear trend, white noise, and red noise produced using a GP. Also plotted is the mean of Gaussian processes consisting of different kernel functions. An ESn2 kernel is shown in blue, the product of an ESn2 kernel and a SE kernel is shown in green, whilst the kernel function described in equation~\protect\ref{kfunc_rot} is shown in red. }
\label{fig:kfuncs}
\end{center}
\end{figure*}
The ability of this Gaussian process to represent quasi-periodic variability with additional systematics is demonstrated in figure~\ref{fig:kfuncs}. We created some fake data by adding together a sinusoidal signal, a longer term linear trend, white noise and red noise produced using a GP with a M32 kernel function. We then calculated the mean of a Gaussian process conditioned on the fake data for three different kernel functions. Figure~\ref{fig:kfuncs} demonstrates that a Gaussian process using a ESn2 kernel function alone can represent periodic variability, but not a long term trend or any evolution in the variability. Multiplying the ESn2 kernel by a SE kernel gives the first term in equation~\ref{kfunc_rot}, and can represent quasi-periodic variability and long term trends. The addition of the M32 kernel allows the Gaussian process to reproduce the quasi-periodic variability and the shorter timescale systematics. Model $\mathcal{M_C}$ is thus capable of faithfully reproducing the lightcurves of variable brown dwarfs which are affected by instrumental systematics or astrophysical red noise (e.g. flaring). 

\subsection{Calculating the marginal likelihoods}

To test if a periodic signal is present in a dataset, we need to be able to calculate the marginal likelihoods for our models $\mathcal{M_A}$, $\mathcal{M_B}$, $\mathcal{M_C}$ along with the uncertainty on the marginal likelihoods. Direct integration of equation~\ref{eq:like} over all of the allowed parameter space is computationally too expensive. \cite{chib01} outline a method whereby the optimal parameters for a model $\mathcal{M}$ can be found via Markov-Chain Monte-Carlo (MCMC) techniques and the marginal likelihood can be estimated from the MCMC chains themselves. This requires a very large number of steps in the MCMC chains, since the marginal likelihood estimate will be dominated by a small fraction of steps 
which lie close to the maximum likelihood. 

Instead, we use parallel-tempering MCMC \cite[see][for a description of the algorithm]{earl05} and calculate the marginal likelihood using thermodynamic integration \citep{goggans04}. We use the parallel tempering algorithm as implemented in {\sc emcee} \citep{foreman-mackey2013}. $N$ parallel MCMC chains are run. Each chain samples from a modified posterior probability given by
\begin{equation}
\pi_{T} (\theta)  = \mathcal{L}(\bm{d} \mid \theta)^{1/T} Pr(\theta),
\end{equation}
where $T$ is the temperature of the chain. For ``hot'' chains, the posterior becomes the prior and the chain explores a wide range of parameter space. Cold chains eplore the peaks of the likelihood function. During the MCMC fit, chains of different temperatures swap parameter values; this helps convergence if the likelihood function is multi-modal. 

We use parallel-tempering MCMC not to improve convergence, but to allow an estimate of the marginal likelihood. We define the inverse temperature $\beta = 1/T$. The marginal likelihood as a function of inverse temperature is
\begin{equation}
m(\beta) = \int{ \mathcal{L} (\mathbfit{d} \mid  \theta)^{\beta} \, Pr(\theta) \; d\theta }.
\end{equation}
 \cite{goggans04} show that the marginal likelihood for a model can be written as:
 \begin{equation}
\ln m = \int_{0}^{1} \langle  \ln \mathcal{L} (\mathbfit{d} \mid  \theta)^{\beta} \rangle  d\beta,
\end{equation}
where $\langle \ln \mathcal{L} (\mathbfit{d} \mid  \theta)^{\beta} \rangle$ is the average log-likelihood of an MCMC chain at inverse temperature $\beta$. This integral can easily be estimated using a quadrature formula from the parallel-tempering MCMC chains, and the uncertainty on the integral arising from using a finite number of temperatures estimated.

\subsection{Tests on simulated data}
\begin{figure}
\begin{center}
\includegraphics[scale=0.4,clip=true,trim=20 0 0 20]{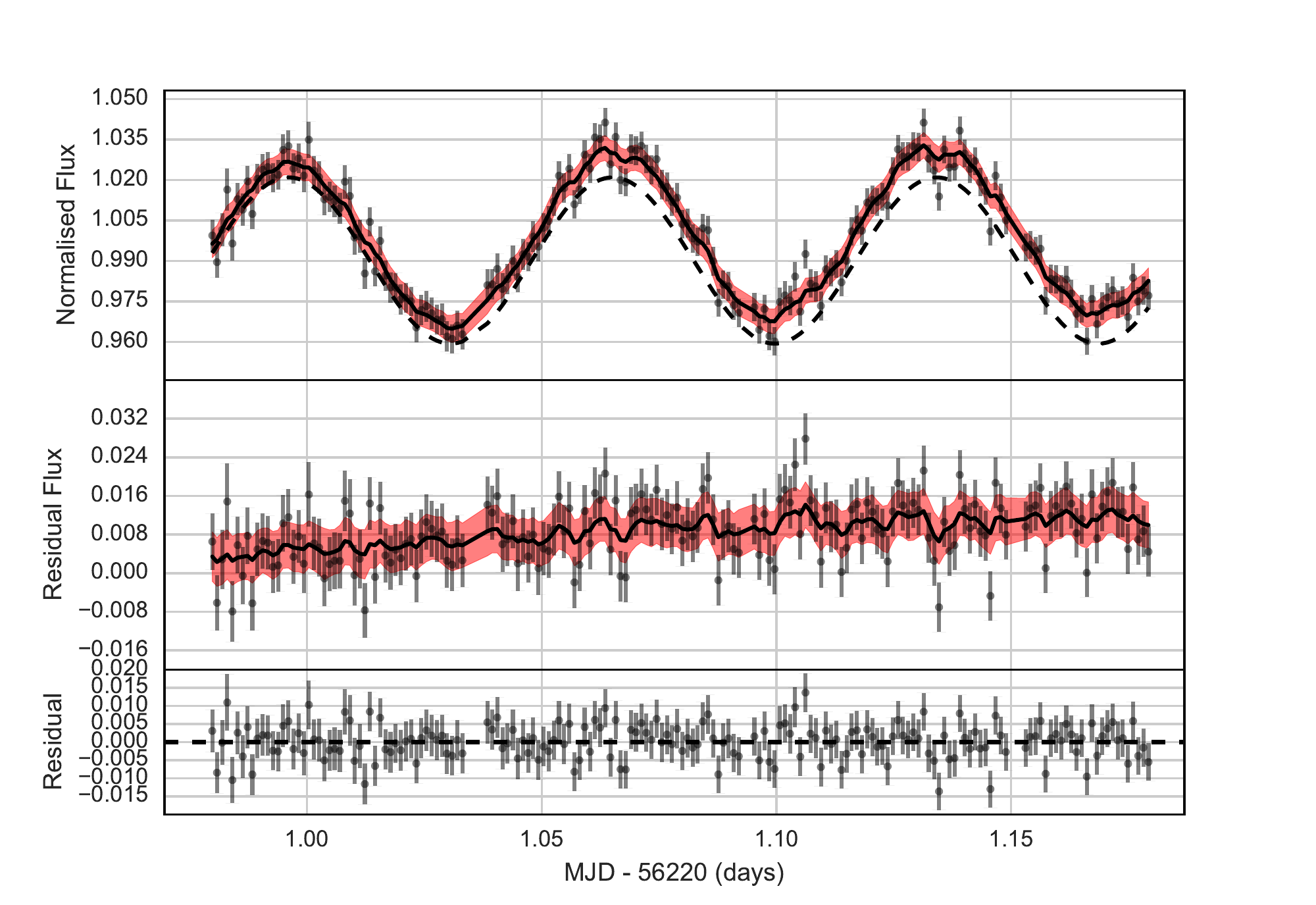}
\includegraphics[scale=0.4,clip=true,trim=20 0 0 20]{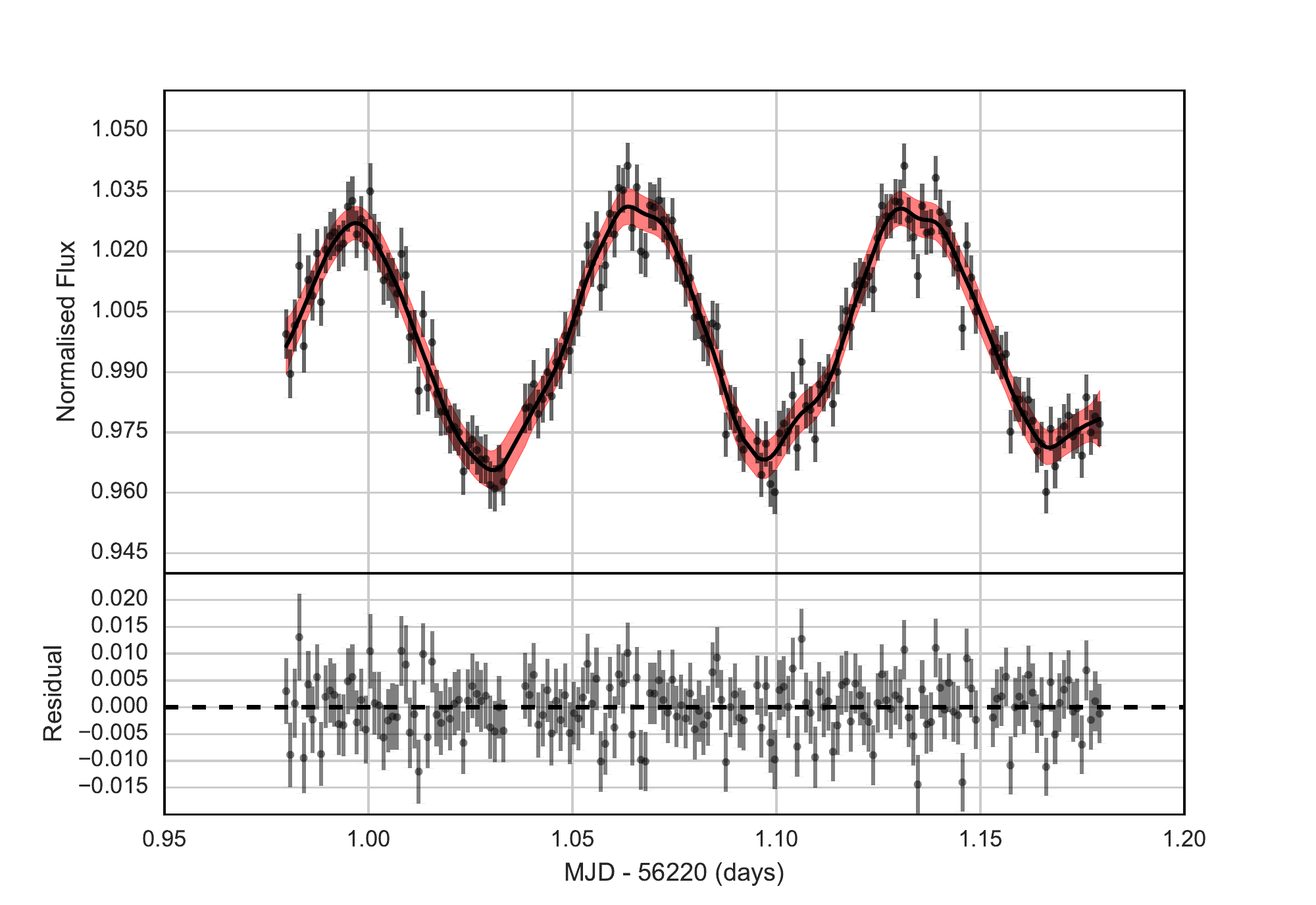}
\includegraphics[scale=0.4,clip=true,trim=20 0 0 20]{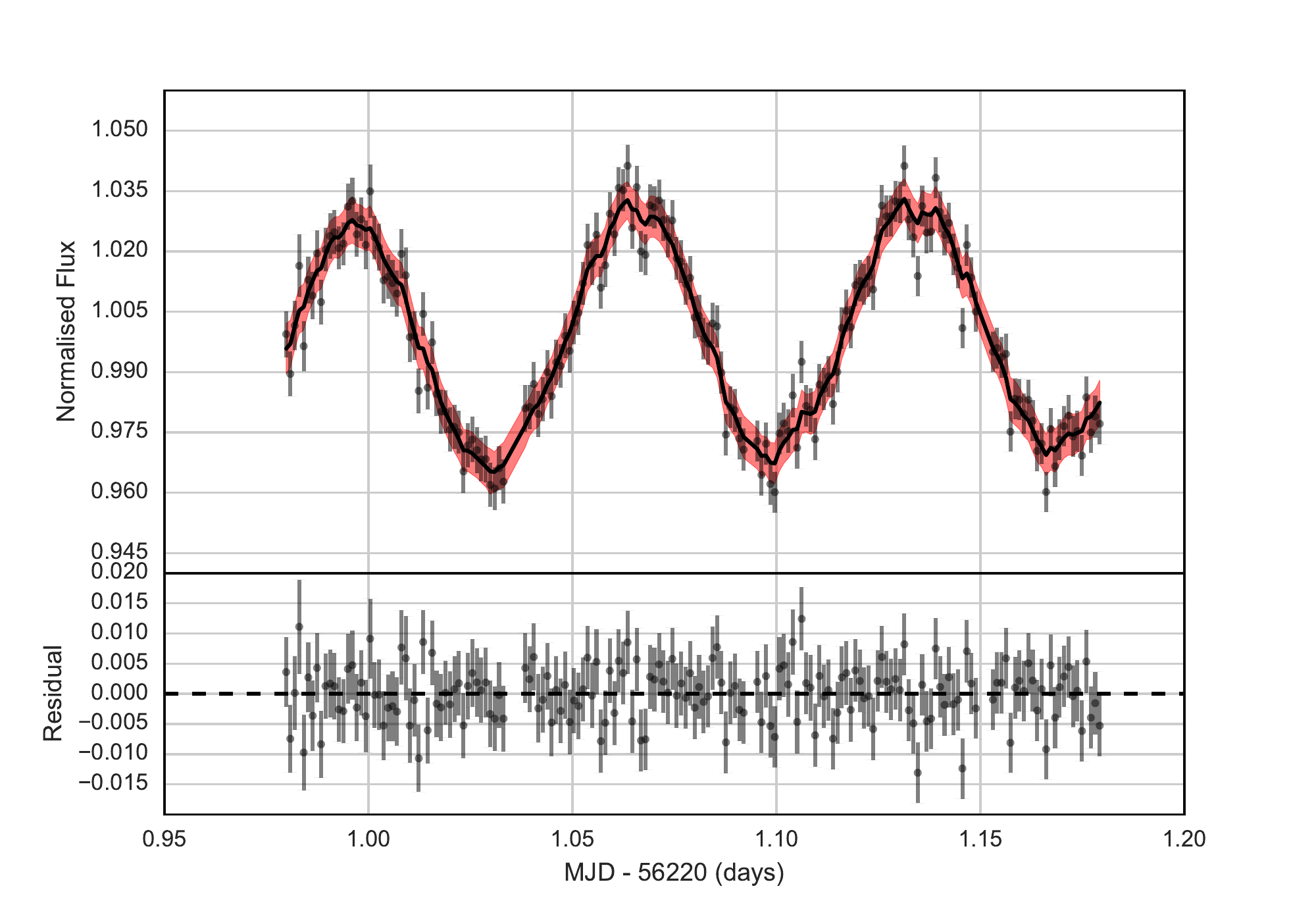}
\caption{Fits to a simulated dataset (see text for details). From top to bottom we show fits using $\mathcal{M_A}$ (red noise plus a sinusoid), $\mathcal{M_B}$ (red noise alone) and $\mathcal{M_C}$ (red noise plus quasi-periodic variability). In each plot the top panel shows the data and the mean (black line) and standard deviation (red region) of the best-fitting model. The bottom panel shows the residuals to the fit. For $\mathcal{M_A}$ (top), a dashed line in the top panel shows the sinusoidal
contribution to the best fit. The middle panel of this plot shows the data and the sinusoidal contribution subtracted, with the
Gaussian process red noise model overplotted.}
\label{fig:fakedata}
\end{center}
\end{figure}
To demonstrate the application of the method to data in which a clear sinusoidal signal is visible but which is also affected by systematics, we added a sinusoidal signal with period 1.66 hours and amplitude of 30 mmag to the lightcurve of ULAS~J2321+1354. We then fit this fake data with models $\mathcal{M_A}$, $\mathcal{M_B}$, $\mathcal{M_C}$ using the parallel-tempering MCMC method described above. Each model was fit using 30 chains of different temperatures. Each chain had an ensemble of 400 walkers to help with convergence and the MCMC was run for 1000 burn-in steps and 1000 production steps. Priors were chosen to be uninformative in most cases.  The rotational period $P$, had a uniform prior which was fixed between 0.5 and 6 hours. To prevent degeneracy between the kernel function terms for short and long timescale variability, $\tau_{l}$ had a log-uniform prior between 2 hours and 1 day whilst $\tau_{s}$ had a log-uniform prior between 1 minute and 1 hour.
We tested that the best fits and marginal likelihood had no significant dependence on the exact choice of priors. There was also no significant change in best fit or marginal likelihood for MCMC runs from different starting positions, or with increased numbers of temperatures, walkers or steps.

The fits to the data are shown in figure~\ref{fig:fakedata}. Once the optimal parameters for the Gaussian process are known, samples can be drawn from the conditional probability for the value of the Gaussian process, given the observed data \citep{rasmussen06}.  The mean and standard deviation of these samples is calculated, and also shown in figure~\ref{fig:fakedata}. The considerable flexibility of GPs to represent time-series data means that reasonable fits are obtained in each case. However, the posterior probability of the best fits are very different for each model. For model $\mathcal{M_A}$, the sinusoidal variability is modelled by the mean function, and the GP parameters are optimised to fit the systematics. For model $\mathcal{M_B}$, the absence of a sinusoidal term means that $a_{l}$ and $\tau_{l}$ are optimised to reproduce the periodic variability and the Gaussian process is not as effective at reproducing the systematics. This makes the posterior probability lower.  This is reflected in the marginal likelihoods of the models. The marginal likelihoods for each model are $\ln m_{\mathcal{A}} = 628 \pm 3$, $\ln m_{\mathcal{B}} = 603 \pm 2$ and $\ln m_{\mathcal{C}} = 618 \pm 4$. Therefore both of the models with a periodic signal are clearly preferred to one with red noise alone, as expected. 

\subsection{ULAS~J2321+1354}
\begin{figure}
\begin{center}
\includegraphics[scale=0.4,clip=true,trim=20 0 0 20]{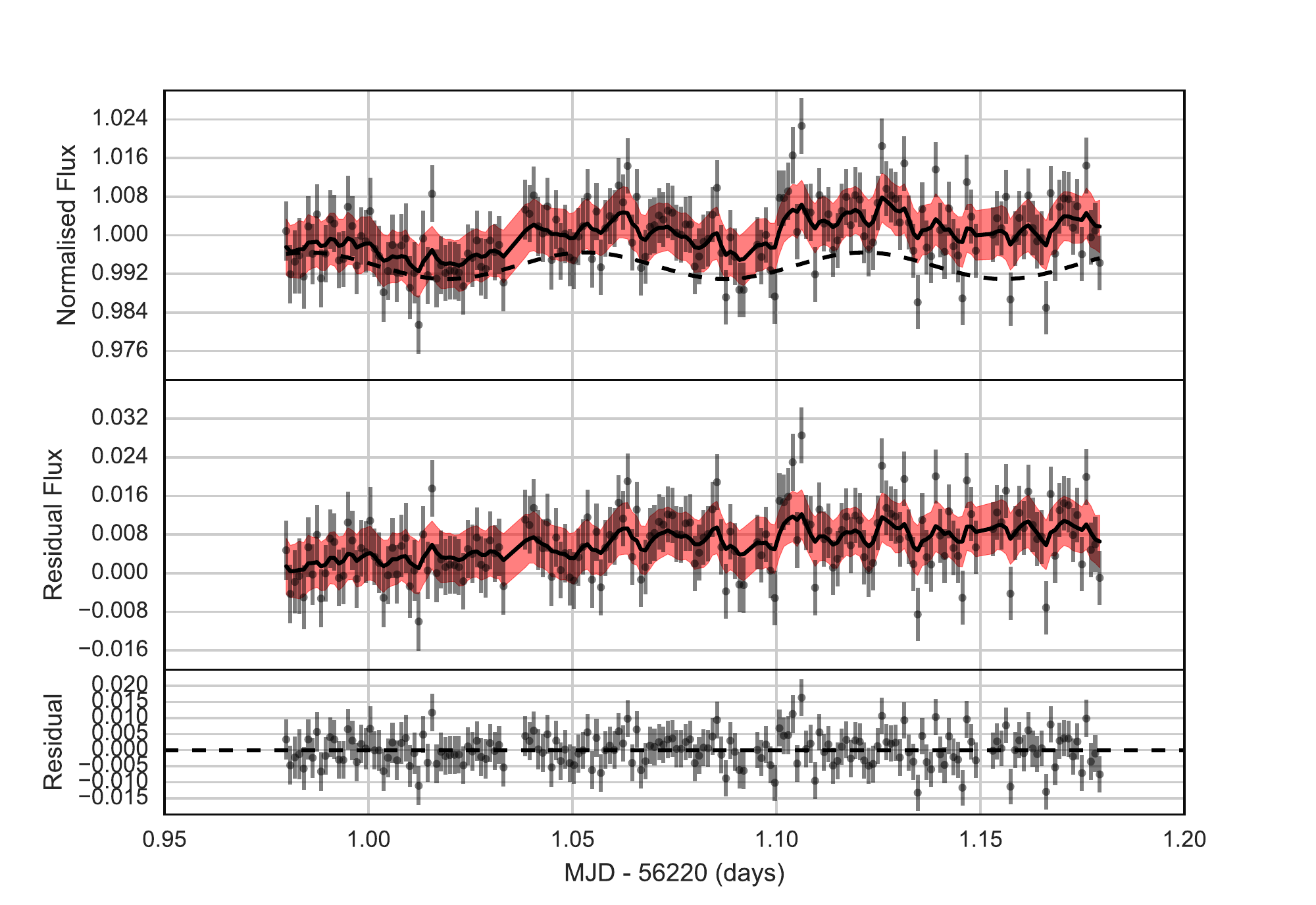}
\includegraphics[scale=0.4,clip=true,trim=20 0 0 20]{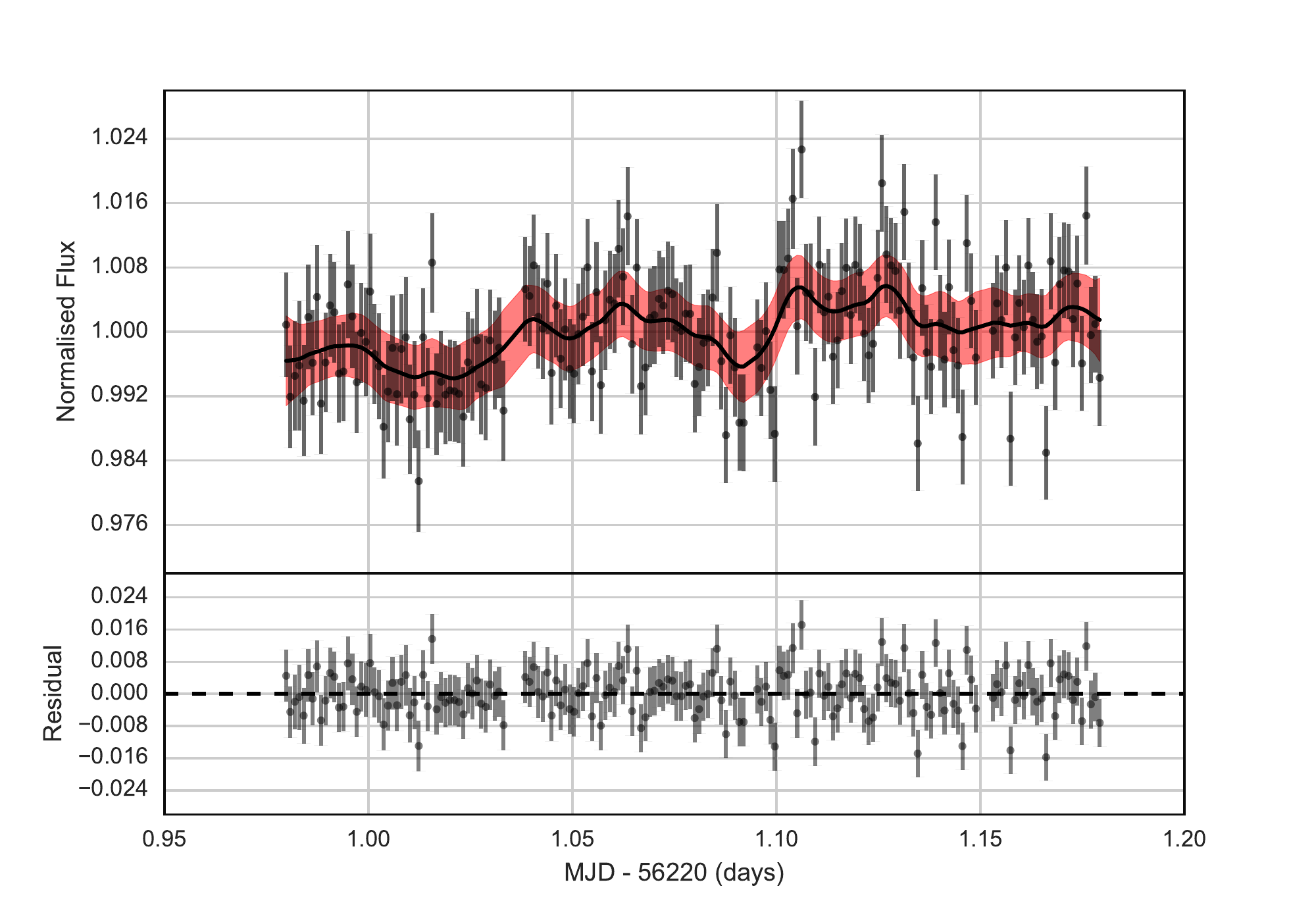}
\includegraphics[scale=0.4,clip=true,trim=20 0 0 20]{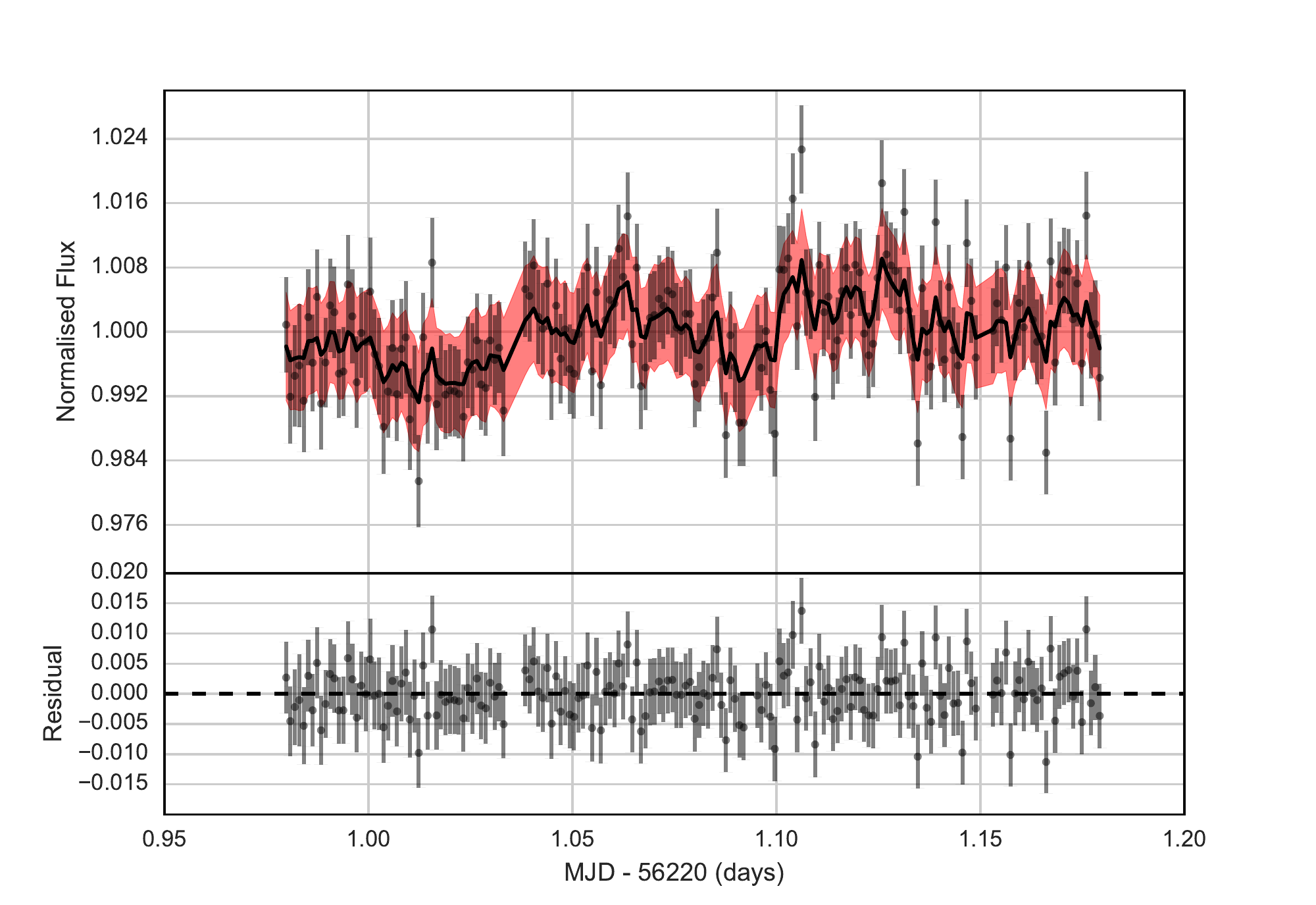}
\caption{Fits the the $J$-band lightcurve of ULAS~J2321+1354. From top to bottom we show fits using $\mathcal{M_A}$ (red noise plus a sinusoid), $\mathcal{M_B}$ (red noise alone) and $\mathcal{M_C}$ (red noise plus quasi-periodic variability). In each plot the top panel shows the data and the mean (black line) and standard deviation (red region) of the best-fitting model. The bottom panel shows the residuals to the fit. For $\mathcal{M_A}$ (top), a dashed line in the top panel shows the sinusoidal
contribution to the best fit. The middle panel of this plot shows the data and the sinusoidal contribution subtracted with the
Gaussian process red noise model overplotted.}
\label{fig:ulas2321}
\end{center}
\end{figure}
We applied the same model fitting process to the $J$-band lightcurve of ULAS~J2321+1354. We do not attempt to remove systematics by detrending before fitting. To do so would mean that uncertainties in the detrending process are not reflected in the final parameter determinations. The fits to the lightcurve for each model, and the conditional distributions of the best fitting Gaussian process, are shown in figure~\ref{fig:ulas2321}. The Bayes' factors for each model show no preference for the models with a periodic component. The marginal likelihoods for each model are $\ln m_{\mathcal{A}} = 621 \pm 3$,  $\ln m_{\mathcal{B}} = 620 \pm 2$ and  $\ln m_{\mathcal{C}} = 621 \pm 3$. 

One might argue that the instrumental systematics affecting the T-dwarfs and reference stars are similar. As discussed in section~\ref{sec:results}, secondary extinction may affect late T-dwarfs differently, but this is an exception: most systematic effects should influence target and reference stars alike. Thus, one may do better than adopting uninformative priors on the red noise terms in our models. Instead, we can fit model $\mathcal{A}$ to our reference lightcurves, and adopt the posterior distribution of parameters as our priors on the matching parameters in models $\mathcal{B}$ and $\mathcal{C}$ when fitting ULAS~J2321+1354. However, this makes little difference to our results. Adopting this approach yields marginal likelihoods for each model of $\ln m_{\mathcal{A}} = 632 \pm 2$,  $\ln m_{\mathcal{B}} = 633 \pm 2$ and  $\ln m_{\mathcal{C}} = 633 \pm 3$.

\subsection{Constraints on periodic variability}
\label{subsec:constraints}
\begin{figure*}
\begin{center}
\includegraphics[scale=0.5,clip=true,trim=20 20 0 0]{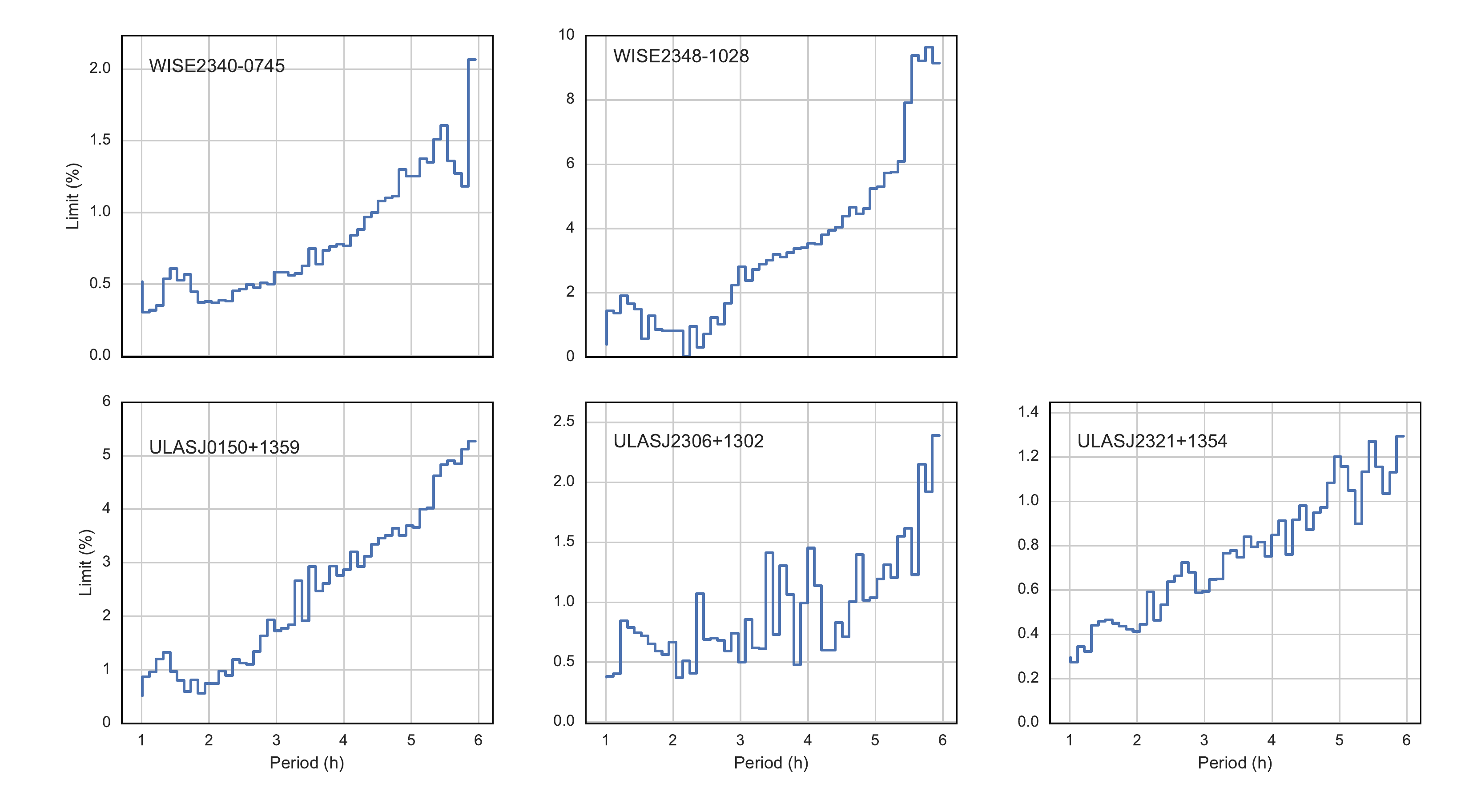}
\caption{Constraints on periodic variability in our target stars. The lines show the 99\% confidence limit to the amplitude of sinusoidal variability, as a function of period, for each star. }
\label{fig:limits}
\end{center}
\end{figure*}
We attempted to place limits on the maximum amplitude of periodic variability in our target stars. To do so, we fit model $\mathcal{B}$ to the un-corrected data for each star and calculated the 99th-percentile limit on the amplitude of the periodic term as a function of period. The results are shown in figure~\ref{fig:limits}. As one might expect, we are able to place tighter limits on the amplitude of periodic variability at shorter periods, and constraints at longer periods are weaker. Typically, we are able to set limits of 0.5--1.0 percent on periodic variability with timescales less than 5 hours. Constraints for ULASJ0150+1359 and WISEJ2348-1020 are weaker due to the larger amplitude of systematics in the lightcurves for these targets.

Since we have detected no evidence for short-term quasi-periodic variability in five targets that were selected on the basis of discrepant $Y-$ and $J-$band photometry on longer timescales, we investigated if secondary extinction could be responsible for the discrepancies seen in the long-term photometry. By folding model atmospheres through the $J$-band filter, after applying telluric absorption using the Mauna Kea transmission profiles of \cite{lord92}, we can estimate the impact of secondary extinction on our measurements for a range of precipitable water columns. We used COND model atmospheres \citep{allard01} with effective temperatures of 1000\,K and 4000\,K to represent our target and a 'typical' reference star respectively. We found that changes in water column from 1 to 5\, mm can introduce spurious variability of up to 0.1 mags if not properly accounted for. As a result, we cannot rule out the possibility that this is the origin of the discrepant $Y-$ and $J-$band photometry used to select our target stars.

\section{Discussion}
\label{sec:discussion}
By comparing the evidence for models with and without periodic signals we see that there is no robust evidence for periodic variability in ULAS~J2321+1354. This is in stark contrast to an analysis using Lomb-Scargle periodograms and the FAP. This is perhaps unsurprising, since the FAP measures the probability a given lightcurve could arise if the data were Gaussian (white) noise and the lightcurve of ULAS~J2321+1354 shows clear red noise which we attribute to instrumental systematics. 

The difficulty of interpreting FAPs in data where any sinusoidal signal is accompanied by red noise (either from instrumental systematics or astrophysical sources) is well known. A commonly adopted solution \cite[see][for example]{littlefair10, metchev15} is to apply a stricter FAP criterion, where the critical FAP is arrived at via some ad-hoc approach. Usually this involves looking at the FAP distribution of stars assumed not to show periodic variability, and choosing a critical FAP high enough to avoid false positive claims. However, this approach is not ideal, since it assumes that the red noise in the control stars used is similar to that in the suspected periodic variables. This may well not be true. Had we adopted this approach for ULAS~J2321+1354 we would have concluded that the periodic variability is real; it has by far the lowest FAP of all of the stars in the field of view.

Our approach using Bayesian model comparison reaches a different conclusion. This is because we are asking if the variability exhibited by ULAS~J2321+1354 is likely to have arisen from systematics alone. Given the presence of both short timescale systematics and a long term systematic trend of similar amplitude to any possible periodic variability, it is not surprising that we find no evidence for a periodic signal in ULAS~J2321+1354. The Bayesian model comparison approach is thus more conservative, and less likely to mistake instrumental systematics for quasi-periodic variability. It is, however, more computationally expensive, with our models taking ${\sim}1$ hour to run on a 64-core desktop machine. We would thus recommend a hybrid approach when dealing with large samples of stars, where candidates for periodic variability are first selected via a Lomb-Scargle periodogram and the reality of that signal is then assessed using model comparison.

\section{Conclusions}
\label{sec:conclusions}
We observed five late T dwarfs with HAWK-I on the VLT for a typical duration of ${\sim}4$ hours. Our data are affected by instrumental systematics with amplitudes between 7 and 30 mmags. Linear detrending for systematics that correlate with sky background, FWHM and airmass removes some, but not all, of these systematics. We are able to place limits on the amplitude of any short periodic variability of 0.5--5 percent, depending on the level of systematics in each lightcurve.

One object, ULAS J2321+13 appears to show periodic variability with a period of 1.64 hours and an amplitude of 3 mmags. A Lomb-Scargle periodogram appears to confirm this variability but a Bayesian comparison of models with and without periodic terms finds no significant evidence that this periodic signal is real. We therefore conclude that none of our targets show periodic variability with amplitudes of a 1.5 percent or greater.

\section{Acknowledgements}
SPL is supported by STFC grant ST/M001350/1. BB acknowledges financial support from the European Commision in the form of a Marie Curie International Outgoing Fellowship (PIOF-GA-2013-629435). Based on observations made with ESO Telescopes at the La Silla Paranal Observatory under programme ID 090.C-0721. ChH highlights financial support of the European Community under the FP7 by the ERC starting grant 257431. This research made use of Astropy, a community-developed core Python package for Astronomy (Astropy Collaboration, 2013). We thank the anonymous referee for their comments, which significantly improved the paper.

\bibliography{refs,refs2,refs3}

\begin{thebibliography}{}
\makeatletter
\relax
\def\mn@urlcharsother{\let\do\@makeother \do\$\do\&\do\#\do\^\do\_\do\%\do\~}
\def\mn@doi{\begingroup\mn@urlcharsother \@ifnextchar [ {\mn@doi@}
  {\mn@doi@[]}}
\def\mn@doi@[#1]#2{\def\@tempa{#1}\ifx\@tempa\@empty \href
  {http://dx.doi.org/#2} {doi:#2}\else \href {http://dx.doi.org/#2} {#1}\fi
  \endgroup}
\def\mn@eprint#1#2{\mn@eprint@#1:#2::\@nil}
\def\mn@eprint@arXiv#1{\href {http://arxiv.org/abs/#1} {{\tt arXiv:#1}}}
\def\mn@eprint@dblp#1{\href {http://dblp.uni-trier.de/rec/bibtex/#1.xml}
  {dblp:#1}}
\def\mn@eprint@#1:#2:#3:#4\@nil{\def\@tempa {#1}\def\@tempb {#2}\def\@tempc
  {#3}\ifx \@tempc \@empty \let \@tempc \@tempb \let \@tempb \@tempa \fi \ifx
  \@tempb \@empty \def\@tempb {arXiv}\fi \@ifundefined
  {mn@eprint@\@tempb}{\@tempb:\@tempc}{\expandafter \expandafter \csname
  mn@eprint@\@tempb\endcsname \expandafter{\@tempc}}}

\bibitem[\protect\citeauthoryear{{Ackerman} \& {Marley}}{{Ackerman} \&
  {Marley}}{2001}]{ackerman01}
{Ackerman} A.~S.,  {Marley} M.~S.,  2001, \mn@doi [\apj] {10.1086/321540},
  \href {http://ukads.nottingham.ac.uk/abs/2001ApJ...556..872A} {556, 872}

\bibitem[\protect\citeauthoryear{{Aigrain}, {Parviainen}  \& {Pope}}{{Aigrain}
  et~al.}{2016}]{aigrain16}
{Aigrain} S.,  {Parviainen} H.,   {Pope} B.~J.~S.,  2016, \mn@doi [\mnras]
  {10.1093/mnras/stw706}, \href
  {http://ukads.nottingham.ac.uk/abs/2016MNRAS.tmp..598A} {}

\bibitem[\protect\citeauthoryear{{Allard}, {Hauschildt}, {Alexander}, {Tamanai}
   \& {Schweitzer}}{{Allard} et~al.}{2001}]{allard01}
{Allard} F.,  {Hauschildt} P.~H.,  {Alexander} D.~R.,  {Tamanai} A.,
  {Schweitzer} A.,  2001, \mn@doi [\apj] {10.1086/321547}, \href
  {http://adsabs.harvard.edu/abs/2001ApJ...556..357A} {556, 357}

\bibitem[\protect\citeauthoryear{{Ambikasaran}, {Foreman-Mackey}, {Greengard},
  {Hogg}  \& {O'Neil}}{{Ambikasaran} et~al.}{2014}]{ambikasaran14}
{Ambikasaran} S.,  {Foreman-Mackey} D.,  {Greengard} L.,  {Hogg} D.~W.,
  {O'Neil} M.,  2014, preprint, \href
  {http://ukads.nottingham.ac.uk/abs/2014arXiv1403.6015A} {} (\mn@eprint
  {arXiv} {1403.6015})

\bibitem[\protect\citeauthoryear{{Artigau}, {Bouchard}, {Doyon}  \&
  {Lafreni{\`e}re}}{{Artigau} et~al.}{2009}]{artigau09}
{Artigau} {\'E}.,  {Bouchard} S.,  {Doyon} R.,   {Lafreni{\`e}re} D.,  2009,
  \mn@doi [\apj] {10.1088/0004-637X/701/2/1534}, \href
  {http://ukads.nottingham.ac.uk/abs/2009ApJ...701.1534A} {701, 1534}

\bibitem[\protect\citeauthoryear{{Bailey}, {Helling}, {Hodos{\'a}n}, {Bilger}
  \& {Stark}}{{Bailey} et~al.}{2014}]{bailey14}
{Bailey} R.~L.,  {Helling} C.,  {Hodos{\'a}n} G.,  {Bilger} C.,   {Stark}
  C.~R.,  2014, \mn@doi [\apj] {10.1088/0004-637X/784/1/43}, \href
  {http://adsabs.harvard.edu/abs/2014ApJ...784...43B} {784, 43}

\bibitem[\protect\citeauthoryear{{Burgasser}, {Marley}, {Ackerman}, {Saumon},
  {Lodders}, {Dahn}, {Harris}  \& {Kirkpatrick}}{{Burgasser}
  et~al.}{2002}]{burgasser02}
{Burgasser} A.~J.,  {Marley} M.~S.,  {Ackerman} A.~S.,  {Saumon} D.,  {Lodders}
  K.,  {Dahn} C.~C.,  {Harris} H.~C.,   {Kirkpatrick} J.~D.,  2002, \mn@doi
  [\apjl] {10.1086/341343}, \href
  {http://ukads.nottingham.ac.uk/abs/2002ApJ...571L.151B} {571, L151}

\bibitem[\protect\citeauthoryear{{Burningham} et~al.,}{{Burningham}
  et~al.}{2010}]{burningham10}
{Burningham} B.,  et~al., 2010, \mn@doi [\mnras]
  {10.1111/j.1365-2966.2010.16800.x}, \href
  {http://ukads.nottingham.ac.uk/abs/2010MNRAS.406.1885B} {406, 1885}

\bibitem[\protect\citeauthoryear{{Burningham} et~al.,}{{Burningham}
  et~al.}{2013}]{burningham13}
{Burningham} B.,  et~al., 2013, \mn@doi [\mnras] {10.1093/mnras/stt740}, \href
  {http://ukads.nottingham.ac.uk/abs/2013MNRAS.433..457B} {433, 457}

\bibitem[\protect\citeauthoryear{{Burrows}, {Sudarsky}  \& {Hubeny}}{{Burrows}
  et~al.}{2006}]{burrows06}
{Burrows} A.,  {Sudarsky} D.,   {Hubeny} I.,  2006, \mn@doi [\apj]
  {10.1086/500293}, \href
  {http://ukads.nottingham.ac.uk/abs/2006ApJ...640.1063B} {640, 1063}

\bibitem[\protect\citeauthoryear{Chib \& Jeliazkov}{Chib \&
  Jeliazkov}{2001}]{chib01}
Chib S.,  Jeliazkov I.,  2001, \mn@doi [Journal of the American Statistical
  Association] {10.1198/016214501750332848}, 96, 270

\bibitem[\protect\citeauthoryear{{Cumming}, {Marcy}  \& {Butler}}{{Cumming}
  et~al.}{1999}]{cumming99}
{Cumming} A.,  {Marcy} G.~W.,   {Butler} R.~P.,  1999, \mn@doi [\apj]
  {10.1086/308020}, \href
  {http://ukads.nottingham.ac.uk/abs/1999ApJ...526..890C} {526, 890}

\bibitem[\protect\citeauthoryear{{Cushing} et~al.,}{{Cushing}
  et~al.}{2016}]{cushing16}
{Cushing} M.~C.,  et~al., 2016, preprint, \href
  {http://ukads.nottingham.ac.uk/abs/2016arXiv160206321C} {} (\mn@eprint
  {arXiv} {1602.06321})

\bibitem[\protect\citeauthoryear{{Earl} \& {Deem}}{{Earl} \&
  {Deem}}{2005}]{earl05}
{Earl} D.~J.,  {Deem} M.~W.,  2005, \mn@doi [Physical Chemistry Chemical
  Physics (Incorporating Faraday Transactions)] {10.1039/b509983h}, \href
  {http://adsabs.harvard.edu/abs/2005PCCP....7.3910E} {7, 3910}

\bibitem[\protect\citeauthoryear{{Foreman-Mackey}, {Hogg}, {Lang}  \&
  {Goodman}}{{Foreman-Mackey} et~al.}{2013}]{foreman-mackey2013}
{Foreman-Mackey} D.,  {Hogg} D.~W.,  {Lang} D.,   {Goodman} J.,  2013, \mn@doi
  [\pasp] {10.1086/670067}, \href
  {http://adsabs.harvard.edu/abs/2013PASP..125..306F} {125, 306}

\bibitem[\protect\citeauthoryear{{Goggans} \& {Chi}}{{Goggans} \&
  {Chi}}{2004}]{goggans04}
{Goggans} P.~M.,  {Chi} Y.,  2004, in {Erickson} G.~J.,  {Zhai} Y.,  eds,
  American Institute of Physics Conference Series Vol. 707, Bayesian Inference
  and Maximum Entropy Methods in Science and Engineering. pp 59--66,
  \mn@doi{10.1063/1.1751356}

\bibitem[\protect\citeauthoryear{{Hallinan} et~al.,}{{Hallinan}
  et~al.}{2015}]{hallinan15}
{Hallinan} G.,  et~al., 2015, \mn@doi [\nat] {10.1038/nature14619}, \href
  {http://ukads.nottingham.ac.uk/abs/2015Natur.523..568H} {523, 568}

\bibitem[\protect\citeauthoryear{{Haywood} et~al.,}{{Haywood}
  et~al.}{2014}]{haywood14}
{Haywood} R.~D.,  et~al., 2014, \mn@doi [\mnras] {10.1093/mnras/stu1320}, \href
  {http://ukads.nottingham.ac.uk/abs/2014MNRAS.443.2517H} {443, 2517}

\bibitem[\protect\citeauthoryear{{Kao}, {Hallinan}, {Pineda}, {Escala},
  {Burgasser}, {Bourke}  \& {Stevenson}}{{Kao} et~al.}{2016}]{kao16}
{Kao} M.~M.,  {Hallinan} G.,  {Pineda} J.~S.,  {Escala} I.,  {Burgasser} A.,
  {Bourke} S.,   {Stevenson} D.,  2016, \mn@doi [\apj]
  {10.3847/0004-637X/818/1/24}, \href
  {http://ukads.nottingham.ac.uk/abs/2016ApJ...818...24K} {818, 24}

\bibitem[\protect\citeauthoryear{{Kissler-Patig} et~al.,}{{Kissler-Patig}
  et~al.}{2008}]{kissler-patig08}
{Kissler-Patig} M.,  et~al., 2008, \mn@doi [\aap]
  {10.1051/0004-6361:200809910}, \href
  {http://esoads.eso.org/abs/2008A%26A...491..941K} {491, 941}

\bibitem[\protect\citeauthoryear{{Line}, {Teske}, {Burningham}, {Fortney}  \&
  {Marley}}{{Line} et~al.}{2015}]{line15}
{Line} M.~R.,  {Teske} J.,  {Burningham} B.,  {Fortney} J.~J.,   {Marley}
  M.~S.,  2015, \mn@doi [\apj] {10.1088/0004-637X/807/2/183}, \href
  {http://ukads.nottingham.ac.uk/abs/2015ApJ...807..183L} {807, 183}

\bibitem[\protect\citeauthoryear{{Littlefair}, {Naylor}, {Mayne}, {Saunders}
  \& {Jeffries}}{{Littlefair} et~al.}{2010}]{littlefair10}
{Littlefair} S.~P.,  {Naylor} T.,  {Mayne} N.~J.,  {Saunders} E.~S.,
  {Jeffries} R.~D.,  2010, \mn@doi [\mnras] {10.1111/j.1365-2966.2010.16066.x},
  \href {http://ukads.nottingham.ac.uk/abs/2010MNRAS.403..545L} {403, 545}

\bibitem[\protect\citeauthoryear{{Lord}, {Hollenbach}, {Colgan}, {Haas},
  {Rubin}, {Erickson}, {Carral}  \& {Maloney}}{{Lord} et~al.}{1992}]{lord92}
{Lord} S.~D.,  {Hollenbach} D.~J.,  {Colgan} S.~W.~J.,  {Haas} M.~R.,  {Rubin}
  R.~H.,  {Erickson} E.~F.,  {Carral} P.,   {Maloney} P.,  1992, in American
  Astronomical Society Meeting Abstracts. p.~1182

\bibitem[\protect\citeauthoryear{{Marley}, {Seager}, {Saumon}, {Lodders},
  {Ackerman}, {Freedman}  \& {Fan}}{{Marley} et~al.}{2002}]{marley02}
{Marley} M.~S.,  {Seager} S.,  {Saumon} D.,  {Lodders} K.,  {Ackerman} A.~S.,
  {Freedman} R.~S.,   {Fan} X.,  2002, \mn@doi [\apj] {10.1086/338800}, \href
  {http://adsabs.harvard.edu/abs/2002ApJ...568..335M} {568, 335}

\bibitem[\protect\citeauthoryear{{Metchev} et~al.,}{{Metchev}
  et~al.}{2015}]{metchev15}
{Metchev} S.~A.,  et~al., 2015, \mn@doi [\apj] {10.1088/0004-637X/799/2/154},
  \href {http://ukads.nottingham.ac.uk/abs/2015ApJ...799..154M} {799, 154}

\bibitem[\protect\citeauthoryear{{Mohanty}, {Basri}, {Shu}, {Allard}  \&
  {Chabrier}}{{Mohanty} et~al.}{2002}]{mohanty02}
{Mohanty} S.,  {Basri} G.,  {Shu} F.,  {Allard} F.,   {Chabrier} G.,  2002,
  \mn@doi [\apj] {10.1086/339911}, \href
  {http://ukads.nottingham.ac.uk/abs/2002ApJ...571..469M} {571, 469}

\bibitem[\protect\citeauthoryear{{Morley}, {Fortney}, {Marley}, {Visscher},
  {Saumon}  \& {Leggett}}{{Morley} et~al.}{2012}]{morley12}
{Morley} C.~V.,  {Fortney} J.~J.,  {Marley} M.~S.,  {Visscher} C.,  {Saumon}
  D.,   {Leggett} S.~K.,  2012, \mn@doi [\apj] {10.1088/0004-637X/756/2/172},
  \href {http://ukads.nottingham.ac.uk/abs/2012ApJ...756..172M} {756, 172}

\bibitem[\protect\citeauthoryear{{Morley}, {Marley}, {Fortney}  \&
  {Lupu}}{{Morley} et~al.}{2014}]{morley14}
{Morley} C.~V.,  {Marley} M.~S.,  {Fortney} J.~J.,   {Lupu} R.,  2014, \mn@doi
  [\apjl] {10.1088/2041-8205/789/1/L14}, \href
  {http://ukads.nottingham.ac.uk/abs/2014ApJ...789L..14M} {789, L14}

\bibitem[\protect\citeauthoryear{{Radigan}}{{Radigan}}{2014b}]{radigan14b}
{Radigan} J.,  2014b, \mn@doi [\apj] {10.1088/0004-637X/797/2/120}, \href
  {http://ukads.nottingham.ac.uk/abs/2014ApJ...797..120R} {797, 120}

\bibitem[\protect\citeauthoryear{{Radigan}, {Jayawardhana}, {Lafreni{\`e}re},
  {Artigau}, {Marley}  \& {Saumon}}{{Radigan} et~al.}{2012}]{radigan12}
{Radigan} J.,  {Jayawardhana} R.,  {Lafreni{\`e}re} D.,  {Artigau} {\'E}.,
  {Marley} M.,   {Saumon} D.,  2012, \mn@doi [\apj]
  {10.1088/0004-637X/750/2/105}, \href
  {http://ukads.nottingham.ac.uk/abs/2012ApJ...750..105R} {750, 105}

\bibitem[\protect\citeauthoryear{{Radigan}, {Lafreni{\`e}re}, {Jayawardhana}
  \& {Artigau}}{{Radigan} et~al.}{2014a}]{radigan14a}
{Radigan} J.,  {Lafreni{\`e}re} D.,  {Jayawardhana} R.,   {Artigau} E.,  2014a,
  \mn@doi [\apj] {10.1088/0004-637X/793/2/75}, \href
  {http://ukads.nottingham.ac.uk/abs/2014ApJ...793...75R} {793, 75}

\bibitem[\protect\citeauthoryear{{Rajan} et~al.,}{{Rajan}
  et~al.}{2015}]{rajan15}
{Rajan} A.,  et~al., 2015, \mn@doi [\mnras] {10.1093/mnras/stv181}, \href
  {http://ukads.nottingham.ac.uk/abs/2015MNRAS.448.3775R} {448, 3775}

\bibitem[\protect\citeauthoryear{Rasmussen \& Williams}{Rasmussen \&
  Williams}{2006}]{rasmussen06}
Rasmussen C.~E.,  Williams C. K.~I.,  2006, {Gaussian processes for machine
  learning}.
Adaptive Computation and Machine Learning, MIT Press, Cambridge, MA, \url
  {http://www.ams.org/mathscinet-getitem?mr=MR2514435}

\bibitem[\protect\citeauthoryear{Roberts, Osborne, Ebden, Reece, Gibson  \&
  Aigrain}{Roberts et~al.}{2013}]{roberts13}
Roberts S.,  Osborne M.,  Ebden M.,  Reece S.,  Gibson N.,   Aigrain S.,  2013,
  \mn@doi [Philosophical Transactions: Mathematical, Physical and Engineering
  Sciences] {10.2307/41739973?ref=no-x-route:3184dfae2e6c38fed03b4a5b05570eed},
  371, 1

\bibitem[\protect\citeauthoryear{{Robinson} \& {Marley}}{{Robinson} \&
  {Marley}}{2014}]{robinson14}
{Robinson} T.~D.,  {Marley} M.~S.,  2014, \mn@doi [\apj]
  {10.1088/0004-637X/785/2/158}, \href
  {http://ukads.nottingham.ac.uk/abs/2014ApJ...785..158R} {785, 158}

\bibitem[\protect\citeauthoryear{{Route} \& {Wolszczan}}{{Route} \&
  {Wolszczan}}{2012}]{route12}
{Route} M.,  {Wolszczan} A.,  2012, \mn@doi [\apjl]
  {10.1088/2041-8205/747/2/L22}, \href
  {http://ukads.nottingham.ac.uk/abs/2012ApJ...747L..22R} {747, L22}

\bibitem[\protect\citeauthoryear{{Route} \& {Wolszczan}}{{Route} \&
  {Wolszczan}}{2016}]{route16}
{Route} M.,  {Wolszczan} A.,  2016, \mn@doi [\apjl]
  {10.3847/2041-8205/821/2/L21}, \href
  {http://ukads.nottingham.ac.uk/abs/2016ApJ...821L..21R} {821, L21}

\bibitem[\protect\citeauthoryear{{Tsuji}, {Ohnaka}, {Aoki}  \&
  {Nakajima}}{{Tsuji} et~al.}{1996}]{tsuji96}
{Tsuji} T.,  {Ohnaka} K.,  {Aoki} W.,   {Nakajima} T.,  1996, \aap, \href
  {http://ukads.nottingham.ac.uk/abs/1996A%26A...308L..29T} {308, L29}

\bibitem[\protect\citeauthoryear{{Vanderburg} et~al.,}{{Vanderburg}
  et~al.}{2015}]{vanderburg15}
{Vanderburg} A.,  et~al., 2015, \mn@doi [\apj] {10.1088/0004-637X/800/1/59},
  \href {http://adsabs.harvard.edu/abs/2015ApJ...800...59V} {800, 59}

\bibitem[\protect\citeauthoryear{Vanderplas}{Vanderplas}{2015}]{vdp15}
Vanderplas J.,  2015, gatspy: General tools for Astronomical Time Series in
  Python, \mn@doi{10.5281/zenodo.14833}, \url
  {http://dx.doi.org/10.5281/zenodo.14833}

\bibitem[\protect\citeauthoryear{{Wilson}, {Rajan}  \& {Patience}}{{Wilson}
  et~al.}{2014}]{wilson14}
{Wilson} P.~A.,  {Rajan} A.,   {Patience} J.,  2014, \mn@doi [\aap]
  {10.1051/0004-6361/201322995}, \href
  {http://ukads.nottingham.ac.uk/abs/2014A%26A...566A.111W} {566, A111}

\bibitem[\protect\citeauthoryear{{Zechmeister} \& {K{\"u}rster}}{{Zechmeister}
  \& {K{\"u}rster}}{2009}]{zechmeister09}
{Zechmeister} M.,  {K{\"u}rster} M.,  2009, \mn@doi [\aap]
  {10.1051/0004-6361:200811296}, \href
  {http://ukads.nottingham.ac.uk/abs/2009A%26A...496..577Z} {496, 577}

\bibitem[\protect\citeauthoryear{{Zhang} \& {Showman}}{{Zhang} \&
  {Showman}}{2014}]{zhang14}
{Zhang} X.,  {Showman} A.~P.,  2014, \mn@doi [\apjl]
  {10.1088/2041-8205/788/1/L6}, \href
  {http://ukads.nottingham.ac.uk/abs/2014ApJ...788L...6Z} {788, L6}

\makeatother
\end{thebibliography}

\end{document}